# Light-matter interaction of single semiconducting AlGaN nanowire and noble metal Au nanoparticle in the sub-diffraction limit

A. K. Sivadasan,[*] Kishore K. Madapu, Sandip Dhara[*]

Nanomaterials and Sensor Section, Surface and Nanoscience Division, Indira Gandhi Centre for Atomic Research, Kalpakkam-60310, India

E-mail: sivankondazhy@gmail.com ; dhara@igcar.gov.in

## Abstract

The near field scanning optical microscopy (NSOM) is not only a tool for imaging of sub-diffraction limited objects but also a prominent characteristic tool for understanding the intrinsic properties of the nanostructures. In order to understand the light-matter interactions in the near field regime using NSOM technique with an excitation of 532 nm (2.33 eV), we selected an isolated single semiconducting AlGaN nanowire (NW) of diameter ~120 nm grown via vapor liquid solid (VLS) mechanism along with metallic Au nanoparticle (NP) catalyst. The role of electronic transitions from different native defect related energy states of AlGaN are discussed in understanding the NSOM images for the semiconducting NW. The effect of strong surface plasmon resonance absorption of excitation laser in the NSOM images for Au NP, involved in the VLS growth mechanism of NWs, is also observed.

Online supplementary data available from (URL)

# 1. Introduction

     The light-matter interaction in the near field regime at the vicinity of nanostructures is a very interesting as well as challenging task in the scientific community. The Abbe's diffraction limit prevents conventional optical microscopes to possess a spatial resolution beyond the value of ~$\lambda/2$, where $\lambda$ is the wavelength of excitation light with a maximum numerical aperture value of unity for the probing objective [1,2]. Thus, even for the visible light of $\lambda=400$ nm cannot image nanostructures of size below 200 nm. The near field scanning optical microscopy (NSOM) assisted with the help of plasmonics is an unique tool to understand the light-matter interaction in the near field regime for optical imaging of nanostructures in the sub-diffraction limit [2-14]. The light passing through the metal coated tip of NSOM probe with a circular aperture diameter of few nanometers at the apex is capable of surpassing the diffraction limit. In the near field regime, the evanescent field emanating from the NSOM probe is not diffraction limited. Hence, it facilitates optical and spectroscopic imaging of objects with nanometer spatial resolution [6,15,16]. The light-matter interaction in metallic nanostructures have opened to a new branch of surface plasmon (SP) based photonics, called plasmonics [17]. The SPs are originated due to the collective oscillation of the free electrons at a frequency equal to the frequency of quantized harmonic oscillation of electrons, $\omega_p$ $(=(n_e e^2/m_{eff}\varepsilon_0)^{½}$, where, $n_e$ is the density, $m_{eff}$ is the effective mass, $e$ is the charge of an electron and $\varepsilon_0$ is the permittivity of free space) about the fixed positive charge centers in the surface of metal nanostructures [6,8-10,16]. The coupling of the incident electromagnetic waves to the coherent oscillation of free-electron plasma near the metal surface is knows as a surface plasmon polariton (SPP) and it is a propagating surface wave at the continuous metal-dielectric interface [18,19]. The electromagnetic field perpendicular to the metal surface decays exponentially and is known as evanescent waves providing sub-wavelength

confinement near to the metal surface [6,17,18]. Matching of the incident excitation frequency ($\omega$) of electromagnetic wave with the plasmon frequency ($\omega_p$) of the electrons in metal nanostructures, leads to an enhanced light-matter interaction, known as surface plasmon resonance (SPR) [6,8,19].

The AlGaN is an intrinsically *n*-type semiconductor and one of the prominent candidate among the group III nitride community with wide, direct and tunable bang gap from 3.4 to 6.2 eV [20,21]. So, the group III nitrides including the ternary alloy of AlGaN nanostructures find tremendous applications in short wavelength and high frequency optoelectronic devices including light emitting diodes, displays and optical communications [22-24].Therefore, by considering the importance of AlGaN nanostructure in the optoelectronic applications as well as semiconducting industries, it is also very important to understand the interaction of AlGaN nanowire (NW) with the visible light.

In the present report, we studied the light-matter interaction of an isolated single and semiconducting AlGaN NW (diameter ~120 nm) grown via vapor liquid solid (VLS) mechanism along with metallic Au nanoparticle (NP) catalysts (diameter ~10-150 nm) in the near field regime by using NSOM technique with a laser excitation of 532 nm (2.33 eV). In order to understand the light-matter interaction of AlGaN NW, we invoked the semiconducting band picture by analysing the different energy states related to native defects with the help of photoluminescence (PL) measurement from a single AlGaN NW. We also studied the effects of plasmonics in the NSOM images of Au NPs which is involved in the VLS growth mechanism of AlGaN NWs.

## 2. Experimental

*2.1 Synthesis and characterization techniques*

The mono-dispersed AlGaN NWs were synthesized by atmospheric pressure chemical vapour deposition (APCVD) technique via VLS growth mechanism using Au NPs as catalysts. The detailed growth process is described in our previous report [25]. The morphological features were studied using a field emission scanning electron microscope (FESEM, SUPRA 55 Zeiss). The structural and crystallographic nature of the AlGaN NW as well as Au-AlGaN NW interface was investigated with the help of a high resolution transmission electron microscopy (HRTEM, LIBRA 200FE Zeiss) by dispersing the NWs in isopropyl alcohol and transferred to Cu grids. To investigate the optical properties, the PL spectra was recorded using a He-Cd UV laser of wavelength 325 nm at room temperature (RT) using a 2400 g.mm$^{-1}$ grating as a monochromatizer and a thermoelectrically cooled CCD detector in the backscattering geometry (inVia; Renishaw, UK). The laser excitations as well as the PL spectra were collected using a 40× micro-spot objective with a numerical aperture (N.A.) value of 0.50.

*2.2 Experimental setup for light-matter interaction in the near field regime*

The NSOM (MultiView 4000; Nanonics, Israel) imaging of nanostructures was used to understand the interaction with 532 nm laser (~ 2.33 eV). The schematic of experimental set up is shown in the figure 1.

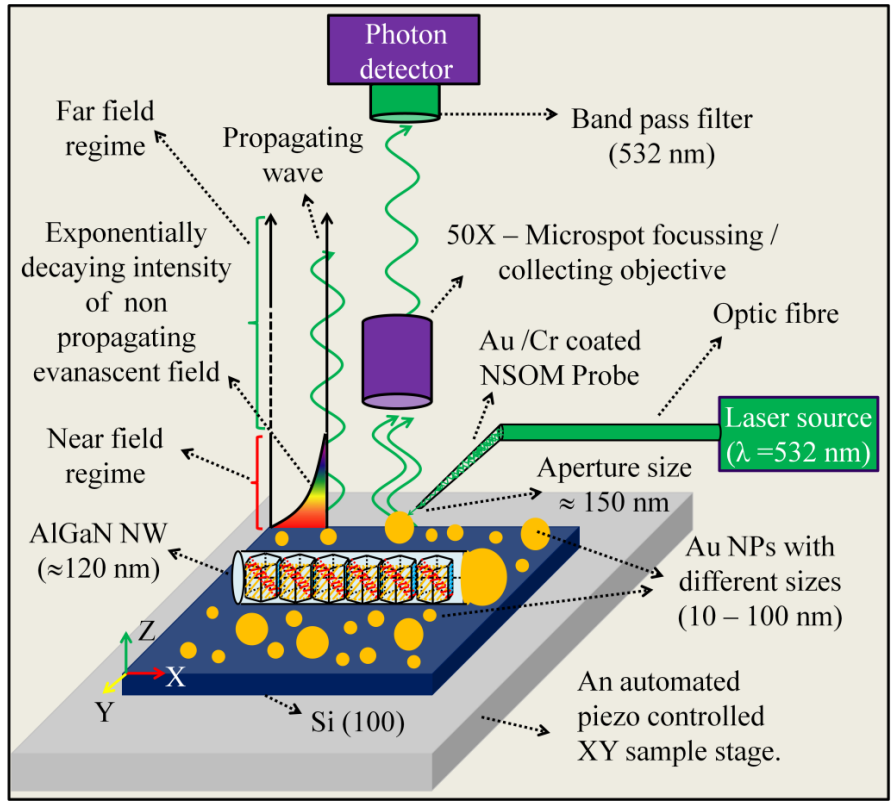

**Figure 1.** The schematic experimental setup for NSOM imaging of AlGaN NW grown via VLS mechanism and Au NPs with the near field excitation and far field collection configuration.

We used an optical fiber with a circular aperture and metal (Au/Cr) coated probe with a tip apex (aperture) diameter of 150 nm for near field excitation of laser light. The optical fiber coated with Cr and Au to avoid leakage of optical power, enhanced optical transmission and confining the light to the sample surface. The scanning was performed either by using the NSOM probe or motorized XY sample stage with very precise special resolution controlled with the help of inbuilt sensors and piezo-drivers. A band pass filter (532 nm) was used to extract the excitation laser after the light-matter interaction and before reaching the light to the detector in the far field configuration. The same probe was used as an atomic force microscopic (AFM) tip for simultaneous scanning of the topography along with the NSOM image of the sample with the tuning fork feedback mechanism.

## 3. Results and discussions

### *3.1 Morphological and structural analysis*

The morphological shape, size and distribution of mono-dispersed AlGaN NWs are shown in the FESEM image (Figure 2). The high resolution FESEM image shows cylindrical shape of the NWs with very smooth surface morphology along with Au catalyst NP at the tip (inset in figure 2). Uniform sized and mono-dispersed NWs with an average diameter of 120 nm were observed. The well separated Au NPs, participated in the VLS growth processes of the NWs, were having uniform size of 150 nm. The Au NPs with diameter varying in the range of 10-75 nm, which did not participate in the growth process, were also found to be distributed over the substrate.

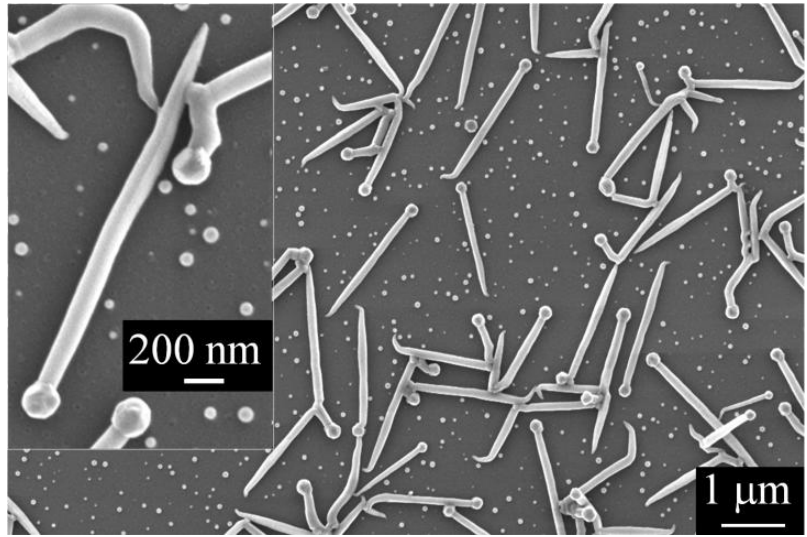

**Figure 2.** The FESEM images of VLS grown mono-dispersed AlGaN NWs with Au NPs as catalysts. Inset shows a typical single AlGaN NW.

The detailed structural analyses of the NWs with nominal presence of Al (discussed in our earlier report),[25] and the Au NP-NW interface studies using high resolution transmission electron microscopy and selected area electron diffraction (SAED).

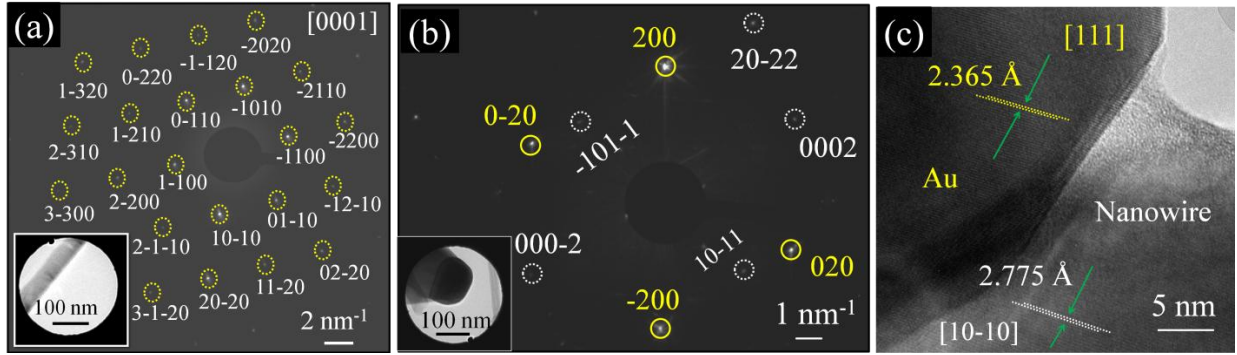

**Figure 3.** (a) The SAED pattern of AlGaN NW with a zone axis of [0001]. The inset showing the portion of AlGaN NW used for the SAED analysis. (b) The SAED pattern of Au-AlGaN NW interface. Spots with the dotted circles indexed to wurtzite GaN and the spots with the circles indexed to cubic Au. The inset showing the Au-AlGaN interface of the NW used for the SAED analysis. (c) The HRTEM image of Au-AlGaN NW interface.

The SAED pattern (Figure 3(a)) of the single NW (inset figure 3(a)) matches well with the wurtzite phase of single crystal GaN with zone axes along [0001]. Moreover, for further understanding of the structural details of Au-AlGaN interface, SAED pattern (Figure 3(b)) of the tip of single AlGaN NW with Au NP at the apex of the NW (inset figure 3(b)) was analysed. The diffraction spots encircled with dotted lines are indexed to wurtzite phase of single crystal AlGaN. Similarly, the diffraction spots encircled with line are indexed to cubic phase of Au. A HRTEM image of Au NP-NW interface is also shown (Figure 3(c)). An inter-planar spacing of 2.365 Å, observed for the NP, corresponds to [111] plane of cubic Au phase. Similarly, in case of NW region, it shows an inter-planar spacing of 2.775 Å corresponding to [10-10] of AlGaN. So the structural analysis confirms presence of AlGaN phase along with the involvement of Au NPs in the VLS growth process of mono-dispersed NWs.

*3.2 Near field light-matter interaction for semiconducting AlGaN NW*

The resultant near field light-matter interaction is shown for AlGaN single NW along with Au NPs of various sizes (Figure 4). The high resolution topographic AFM image of the single NW in the two-dimension (2D) and 3D (Figures 4(a) and 4(b)) shows smooth and cylindrical shape, as observed in the FESEM images, with a diameter of ~ 120 nm. The height variations of AFM cantilever along the line across Au NP-AlGaN NW-Au NP show (Figure 4(c)) the height of AlGaN NW and Au NP as 120 and 40 nm, respectively. The NSOM images of AlGaN single NW as well as catalyst Au NPs in 2D and 3D are also shown (Figures 4(d) and 4(e)). The NSOM imaging for an ensemble of AlGaN NWs shows similar results as observed for the single NW (included in the supplemental materials figure S1 available at URL).

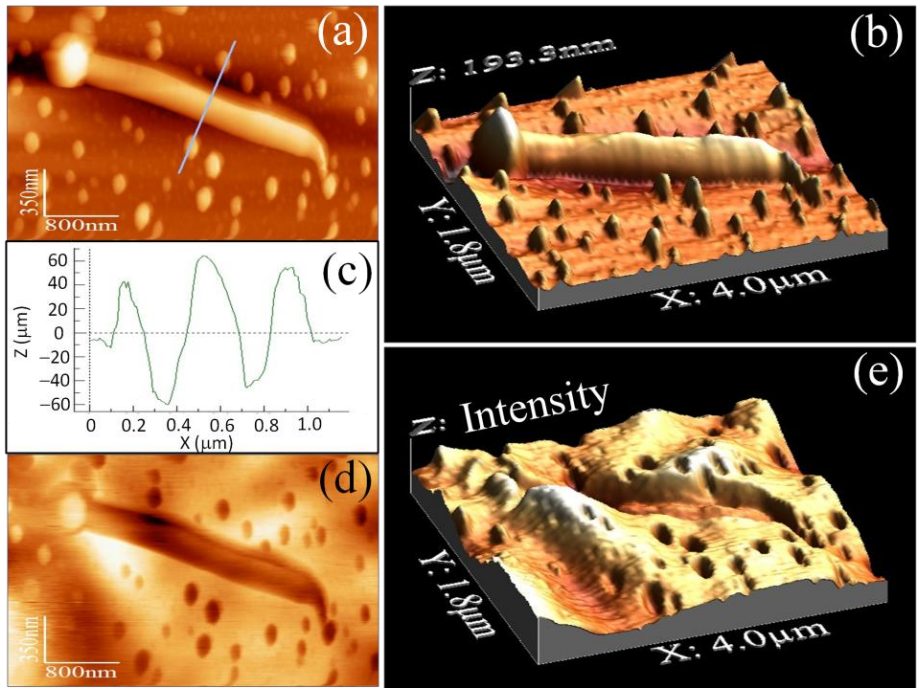

**Figure 4.** The topographic (a) 2D and (b) 3D AFM images of a single AlGaN NW along with Au NPs, (c) The height variation of AFM cantilever along the line across Au NP-AlGaN NW-Au NP as shown in (a) and the corresponding (d) 2D and (e) 3D optical NSOM images of a single AlGaN NW along with Au NPs.

Since the diameter of AlGaN NW (~ 120 nm) is far below the diffraction limit for the excitation wavelength (532 nm), one need to shorten the wavelength down to the sub-diffraction regime for obtaining highly resolved optical image. Using metallic coated NSOM probe, it is possible to produce an evanescent wave with momentum higher than that of the original excitation wavelength $\lambda_0=2\pi/k_0(\omega)$ with wave vector of $k_0(\omega)=\omega/c$, where $c$ is the velocity of light. Therefore, the evanescent waves emanating from the NSOM probe aperture possess group of wave vectors higher than the original excitation laser as $k_{ev}(\omega)=\omega/v$, with different velocities ($v$) slower than the excitation wave velocity ($v < c$). Therefore, the NSOM measurements are advantageous to provide super-resolution. At the same time, it preserves the excitation frequency and energy. Thus, it offers the possibility of optical as well as spectroscopic imaging in the sub-diffraction regime. So, from the scanned images using NSOM technique, we can understand even the intrinsic properties of a sample as revealed by its electronic or vibrational energies [2]. By taking this advantage, we used the NSOM technique to understand the interaction of green laser (532 nm = 2.33 eV) with single semiconducting AlGaN NW in the near-field regime. The reported RT band gap of our AlGaN NWs is 3.53 eV,[25] which is higher than the excitation energy of 2.33 eV. Therefore, a complete transmission of light through the AlGaN NW is expected. Surprisingly, however we observed a prominent absorption of light along the AlGaN NW, as shown in the NSOM images in 2D and 3D (Figures 4(d) and 4(e)).

For better understanding of the light-matter interaction of AlGaN NW, we recorded PL spectrum (Figure 5) in the energy range of 1.7 to 3.85 eV from an ensemble and single NW using an UV excitation of 325 nm (~3.85 eV). Distinct PL peaks centered at 3.53, 3.30, 2.05, and 1.76 eV are observed (Figure 5).

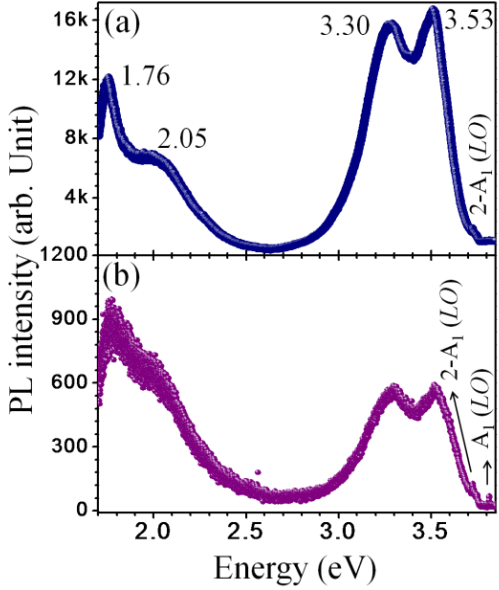

**Figure 5.** The room temperature PL spectrum for (a) an ensemble and (b) a single AlGaN NW with an excitation 325 nm (3.85 eV).

The PL spectrum, recorded for an ensemble as well as single NW using an available visible excitation of 514.5 nm (~2.41 eV) which is close to the excitation used in NSOM measurement, also shows the similar low energy emission peaks centered at ~2.08, and ~1.77 eV (Figure 6(a)). The PL emission at 3.53 eV is reported to originate because of the recombination of free exciton (FE) from the conduction band minimum to the valence band maximum,[25] as shown in the schematic band-diagram (Figure 6(b)). The emission peak observed at 3.30 eV is because of the recombination of the neutral donor-acceptor pair (DAP; $D^0A^0$), with nitrogen vacancy ($V_N$) as the shallow donor and Ga vacancy ($V_{Ga}$) as a deep acceptor state [25]. The presence of unintentional O dopant in the AlGaN NW may be responsible for the emission at ~2.05 eV (Figure 6(b)), which may be observed because of the recombination of electrons from the shallow donor state of $V_N$ to the holes in the deep acceptor state of $O_N$-$V_{Ga}$ complex [26]. Similarly, the presence of unavoidable impurity of C, in the sample, produces the $C_N$-$V_{Ga}$ complex which also may act as a deep acceptor leading to the PL emission peak centered at 1.76 eV (Figure 6(b)) for recombination of electrons from the shallow donor state of $V_N$, [27,28]. The peaks observed at 3.78 and 3.73 eV are assigned to multi-phonon modes of first order, $A_1$(LO) and corresponding second order $2A_1$(LO) Raman modes of wurtzite AlGaN, respectively [1,25].

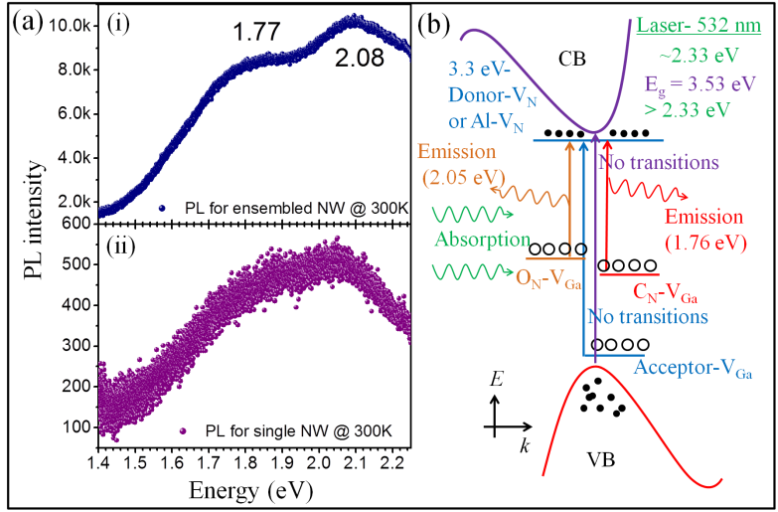

**Figure 6.** (a) The room temperature PL spectrum for (i) an ensemble and (ii) a single AlGaN NW with an excitation of 514.5 nm (2.42 eV). (b) The schematic band diagram of AlGaN NW showing possible light-matter interactions and electronic transitions for 532 nm (2.33 eV) excitation.

The interaction of single AlGaN NW with light in the near-field regime is realized by using the help of possible energy levels in the schematic representation of band-diagram (Figure 6(b)). Even though, we are using the evanescent wave to image the NW, the energy of the electromagnetic field is preserved as 2.33 eV. Therefore, it does not invoke any electronic transition for DAP and FE emission at 3.30 and 3.53 eV, respectively, which are well above 2.33 eV. However, the excitation energy of 2.33 eV is sufficient to activate the transitions at 2.05 and 1.76 eV. The emissions at 1.76 and 2.05 eV are blocked by the band pass filter used before the detector in the NSOM experimental set up (Figure 1). Therefore, the probing light energy is utilized for exciting the defect levels and hence we observe a significant absorption along the single NW in the NSOM images (Figures 4(d) and 4(e)). In the same images, we also observed strong SPR related absorption for Au NPs. Therefore, we carried out NSOM imaging of Au NP for detailed understanding of plasmonic effects, as discussed below.

*3.3 Near field light-matter interaction for metallic Au NP*

We also studied the near field interaction of metallic Au NP with the same visible laser light of energy 2.33 eV (Figure 6). The high resolution topographic 2D and 3D AFM images of the Au NP (Figures 7(a) and 7(b)) shows smooth and spherical shape of the NP with a diameter of ~ 100 nm. The resultant NSOM images of catalyst Au NP in 2D and 3D are shown in figures 7(c) and 7(d).

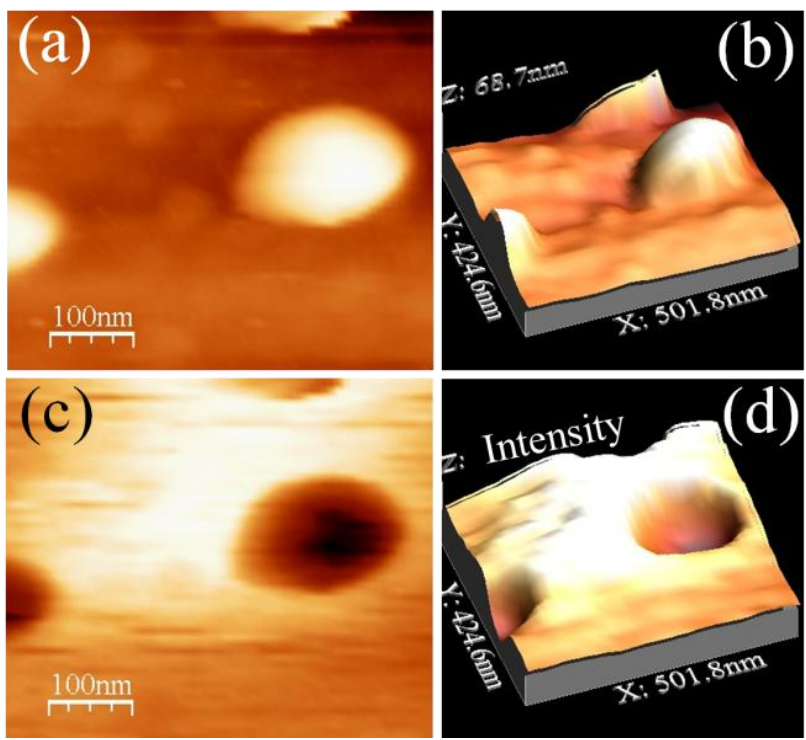

**Figure 7.** (a) 2D and (b) 3D topographic AFM and corresponding (a) 2D and (b) 3D optical NSOM images of a typical Au NP.

The NSOM image of Au NP (Figures 7(c) and 7(d)) also shows strong absorption of electromagnetic waves, but the mechanism behind the light-matter interaction is entirely different from the previous case as in the semiconducting AlGaN NW. However, a significant absorption of light with wavelength 532 nm by Au NP is because of the fact that the SPR peak value ~550 nm for Au NPs matches the excitation wavelength [8,29,30]. The NSOM measurement is also performed for an ensemble of Au NPs showing similar strong SPR absorption (included in the supplemental materials figure S2 available at URL). At resonance, the incident electromagnetic waves coupled with collective oscillation electrons can produce SPP which is perpendicular to the surface of Au NP. The frequency dependent wave vector of SPP can be expressed as in terms of frequency dependent dielectric constants of metal ($\varepsilon_m = \varepsilon_m' + i\varepsilon_m''$) and dielectric ($\varepsilon_d = 1$, for air or vacuum), $k_{spp}(\omega) = \frac{\omega}{c}\sqrt{\frac{\varepsilon_d \cdot \varepsilon_m}{\varepsilon_d + \varepsilon_m}}$. Therefore, the effective wavelength of the SPP is $\lambda_{spp} = 2\pi/k_{spp}$ [7,10,16,31]. The SPPs with a specific wavelength smaller than the excitation can propagate through the surface of Au NPs up to a propagation length which depends on the complex dielectric constants of the metal and dielectric medium [6,31]. Once the SPP propagates through the surface of Au NP and crosses the metallic region, then the electromagnetic wave may decouple from the SPP and it can be converted to a propagating wave. The intensity of the absorption is influenced by the frequency dependent poalrizability of the Au NPs and it can be vary with respect to the size of the Au NPs [4,29-31]. Thus, because of the variation of different sizes of the Au NPs, it is possible to observe them with relatively different absorption intensities (Figures 4(d) and 4(e)). Apart from the formation of SPP, some portion of the absorbed excitation laser may also participate in the lattice phonon generations leading to heating as well as inter-band transitions of Au NPs [8,32].

## 4. Conclusions

We envisage the use of the NSOM technique for understanding the intrinsic characteristics of the nanostructures along with the imaging of sub-diffraction limited objects. The major achievement of our study is the direct understanding of light-matter interaction of semiconducting as well as metallic nanostructures in the near field regime. The isolated single and semiconducting AlGaN nanowire with a diameter ~ 120 nm grown via VLS mechanism shows a strong absorption of visible light due to the electronic transitions originated from the native defect related energy levels. The NSOM images of metallic Au nanoparticle catalyst with diameter ~100 nm shows a strong surface plasmon resonance related absorption of excitation laser with an energy of 2.33 eV ($\lambda$ = 532 nm) due to the formation of surface plasmon polaritons on the surface of the Au nanoparticles.

## Acknowledgments

One of us (AKS) acknowledges the Department of Atomic Energy (DAE), India for permitting him to continue the research work. We thank S. R. Polaki, SND, IGCAR for his help in the FESEM studies. We also acknowledge S. Amirthapandian, Materials Physics Division, IGCAR for structural studies. We also extend our thanks to A. Patsha, S. Parida and R. Basu of MSG, IGCAR for their valuable suggestions and useful discussions.

# References

[1] Sivadasan A K, Patsha A and Dhara S 2015 Optically confined polarized resonance Raman studies in identifying crystalline orientation of sub-diffraction limited AlGaN nanostructure *Appl. Phy. Lett.* **106** 173107

[2] Kawata S, Inouye Y and Verma P 2009 Plasmonics for near-field nano-imaging and superlensing *Nature Photonics* **3** 388

[3] Achermann M 2010 Exciton− plasmon interactions in metal− semiconductor nanostructures *J. Phys. Chem. Lett.* **1** 2837

[4] Rotenberg N and Kuipers L 2014 Mapping nanoscale light fields *Nature Photonics* **8** 919

[5] Okamoto H, Narushima T, Nishiyama Y and Imura K 2015 Local optical responses of plasmon resonances visualised by near-field optical imaging *Phys. Chem. Chem. Phys.* **17** 6192

[6] Barnes W L, Dereux A and Ebbesen T W 2003 Surface plasmon subwavelength optics *Nature* **424** 824

[7] Genet C and Ebbesen T 2007 Light in tiny holes *Nature* **445** 39

[8] Li M, Cushing S K and Wu N 2015 Plasmon-enhanced optical sensors: a review *Analyst* **140** 386

[9] Schuller J A, Barnard E S, Cai W, Jun Y C, White J S and Brongersma M L 2010 Plasmonics for extreme light concentration and manipulation *Nat. Mater.* **9** 193

[10] Weiner J 2009 The physics of light transmission through subwavelength apertures and aperture arrays *Rep. Prog. Phys.* **72** 064401

[11] Lezec H J, Degiron A, Devaux E, Linke R, Martin-Moreno L and Garcia-Vidal F, Ebbesen T 2002 Beaming light from a subwavelength aperture *Science* **297** 820

[12] Ringe E, Sharma B, Henry A I, Marks L D and Van Duyne R P 2013 Single nanoparticle plasmonics *Phys. Chem. Chem. Phys.* **15** 4110

[13] Liu J, Perkins N, Horton M, Redwing J, Tischler M and Kuech T 1996 A near-field scanning optical microscopy study of the photoluminescence from GaN films *Appl. Phys. Lett.* **69** 3519

[14] Hecht B, Sick B, Wild U P, Deckert V, Zenobi R, Martin O J, Pohl D W 2000 Scanning near-field optical microscopy with aperture probes: Fundamentals and applications *J. Chem. Phys.* **112** 7761

[15] Marinello F, Schiavuta P, Cavalli R, Pezzuolo A, Carmignato S and Savio E 2014 Critical Factors in Cantilever Near-Field Scanning Optical Microscopy *Sensors Journal, IEEE* **14** 3236

[16] Zhang J, Zhang L and Xu W 2012 Surface plasmon polaritons: physics and applications *J. Phys. D: Appl. Phys.* **45** 113001

[17] Ozbay E 2006 Plasmonics: merging photonics and electronics at nanoscale dimensions *Science* **311** 189

[18] Kauranen M and Zayats A V 2012 Nonlinear plasmonics *Nature Photonics* **6** 737

[19] Polman A 2008 Plasmonics applied *Science* **322** 868

[20] Pierret A, Bougerol C, Murcia-Mascaros S, Cros A, Renevier H, Gayral B and Daudin B 2013 Growth, structural and optical properties of AlGaN nanowires in the whole composition range *Nanotechnology* **24** 115704

[21] He C, Wu Q, Wang X, Zhang Y, Yang L, Liu N, Zhao Y, Lu Y and Hu Z 2011 Growth and Characterization of Ternary AlGaN Alloy Nanocones across the Entire Composition Range *ACS Nano* **5** 1291

[22]  Pauzauskie P J and Yang P 2006 Nanowire photonics *Materials Today* **9** 36

[23]  Wang Q, Connie A, Nguyen H, Kibria M, Zhao S, Sharif S, Shih I and Mi 2013 Highly efficient, spectrally pure 340 nm ultraviolet emission from $Al_xGa_{1-x}N$ nanowire based light emitting diodes *Nanotechnology* **24** 345201

[24]  Baek H, Lee C H, Chung K and Yi G C 2013 Epitaxial GaN microdisk lasers grown on graphene microdots *Nano. Lett.* **13** 2782

[25]  Sivadasan A K, Patsha A, Polaki S, Amirthapandian S, Dhara S, Bhattacharya A, Panigrahi B K and Tyagi A K 2015 Optical Properties of Monodispersed AlGaN Nanowires in the Single-Prong Growth Mechanism *Cryst. Growth & Des.* **15** 1311

[26]  Patsha A, Amirthapandian S, Pandian R and Dhara S, 2013 Influence of oxygen in architecting large scale nonpolar GaN nanowires *J. Mater. Chem. C.* **1** 8086

[27]  Zhou D, Ni Y, He Z, Yang F, Yao Y, Shen Z, Zhong J, Zhou G, Zheng Y and He L 2016 Investigation of breakdown properties in the carbon doped GaN by photoluminescence analysis *Phys. Status Solidi C*, **1**

[28]  Seager C, Wright A, Yu J and Götz W 2002 Role of carbon in GaN *J. Appl. Phys.* **92** 6553

[29]  Myroshnychenko V, Rodríguez-Fernández J, Pastoriza-Santos I, Funston A M, Novo C, Mulvaney P, Liz-Marzán L M and de Abajo F J G 2008 Modelling the optical response of gold nanoparticles *Chem. Soc. Rev.* **37** 1792

[30]  Maier S A and Atwater H A 2005 Plasmonics: Localization and guiding of electromagnetic energy in metal/dielectric structures *J. Appl. Phys.* **98** 011101

[31]  Murray W A and Barnes W L 2007 Plasmonic materials *Adv. Mater.* **19** 3771

[32]   Beversluis M R, Bouhelier A and Novotny L 2003 Continuum generation from single gold nanostructures through near-field mediated intraband transitions *Phys. Rev. B.* **68** 115433

# Supplemental materials

**Light-matter interaction of single semiconducting AlGaN nanowire and noble metal Au nanoparticle in the sub-diffraction limit**

A. K. Sivadasan,[*] Kishore K. Madapu, Sandip Dhara[*]

Nanomaterials and Sensor Section, Surface and Nanoscience Division, Indira Gandhi Centre for Atomic Research, Kalpakkam-60310, India

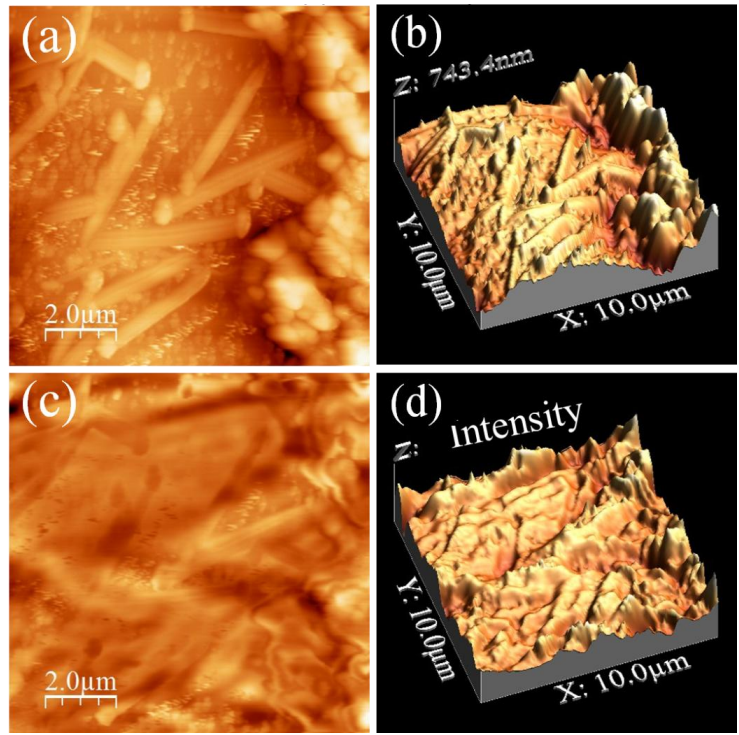

**Figure S1.** The topographic (a) 2D and (b) 3D AFM and the corresponding (c) 2D and (d) 3D optical NSOM images of an ensemble of AlGaN NWs as well as Au NPs participated in the VLS growth process.

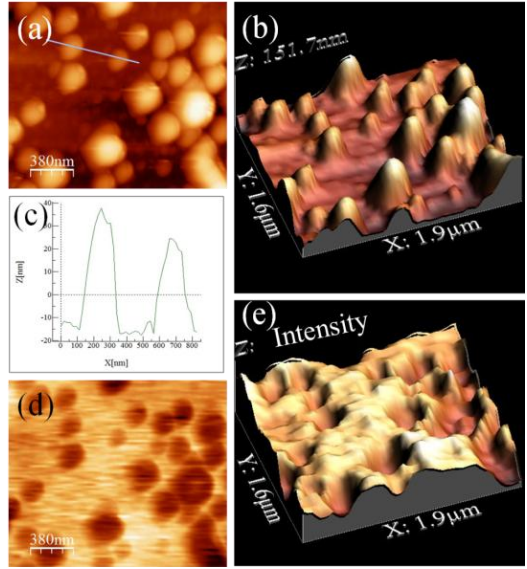

**Figure S2.** The topographic (a) 2D and (b) 3D AFM images of an ensemble of Au NPs. (c) The height variation of AFM cantilever along the line across two Au NPs as shown in (a) and the corresponding (d) 2D and (e) 3D optical NSOM images of an ensemble of Au NPs.